\documentclass[conference]{IEEEtran}
\IEEEoverridecommandlockouts
\usepackage{cite}
\usepackage{amsmath,amssymb,amsfonts}
\usepackage{algorithmic}
\usepackage{graphicx}
\usepackage{textcomp}
\usepackage{xcolor}
\usepackage{subcaption}
\usepackage{comment}
\usepackage{url}
\usepackage{hyperref}

\def\BibTeX{{\rm B\kern-.05em{\sc i\kern-.025em b}\kern-.08em
    T\kern-.1667em\lower.7ex\hbox{E}\kern-.125emX}}
\begin{document}

\title{Emotion-robust EEG Classification for Motor-imagery \\
}

\author{\IEEEauthorblockN{Abdul Moeed}
\IEEEauthorblockA{
\textit{Technical University of Munich}\\
Munich, Germany \\
abd.moeed@tum.de}
}

\maketitle

\begin{abstract}
Developments in Brain Computer Interfaces (BCIs) are empowering those with severe physical afflictions through their use in assistive systems. Common methods of achieving this is via Motor-Imagery (MI), which maps brain signals to code for certain commands. Electroencephalogram (EEG) is preferred for recording brain signal data on account of it being non-invasive. Despite their potential utility, MI-BCI systems are yet confined to research labs. A major cause for this is lack of robustness of such systems. As hypothesized by two teams during Cybathlon 2016, a particular source of the system's vulnerability is the sharp change in the subject's state of emotional arousal. This work aims towards making MI-BCI systems resilient to such emotional perturbations. To do so, subjects are exposed to high and low arousal-inducing virtual reality (VR) environments before recording EEG data. The advent of COVID-19 compelled us to modify our methodology. Instead of training machine learning algorithms to classify emotional arousal, we opt for classifying subjects that serve as proxy for each state. Additionally, MI models are trained for each subject instead of each arousal state. As training subjects to use MI-BCI can be an arduous and time-consuming process, reducing this variability and increasing robustness can considerably accelerate the acceptance and adoption of assistive technologies powered by BCI.
\end{abstract}

\section{Introduction}
Biological systems, such as humans, use electrical signals as the medium of communication between their control centers (brains) and motor organs (arms, legs). While this is taken for granted by most people, those with severe physical impairments, such as quadriplegia, experience the breakdown of this communication system rendering them unable to perform the most basic physical movements. Modern technologies, such as BCIs, have attempted to ameliorate this through the use of brain signals as commands for assistive systems \cite{perdikis2018cybathlon}. MI, a common paradigm for BCI control, requires the subject to simulate or imagine movement of the limbs on account of there being discernible differences in brain signals when moving different limbs \cite{abiri2019comprehensive}. Due to it being non-invasive and cost-effective, EEG is the method of choice for collecting data for such systems \cite{abiri2019comprehensive}.

One of the many recent developments in the application of EEG-driven BCIs is the Cybathlon competition held every four years under the auspices of Eidgenössische Technische Hochschule Zürich (ETH Zurich) \cite{riener2014cybathlon}. The competition involves physically challenged individuals completing routine tasks via assistive systems. One such task -- the BCI race -- has the participants (called pilots) control a virtual game character via brain signals only. Competing teams, who may hail from either academia or industry, are responsible for creating BCI systems and training their respective pilots. The goal of the Cybathlon is to push the state-of-the-art in BCI assistive systems, and accelerate its adoption in everyday lives of those who need it most.

For the 2020 edition of Cybathlon, a team from the Technische Universität München (TUM) called "CyberTUM" is amongst the competitors in the BCI race challenge. In order to achieve high scores in the competition, a major part of BCI development is the focus on robustness of the system i.e. minimizing the variability of the system for different sessions and environments. Lack of robustness, in fact, is an established concern in almost all BCI systems. Possible causes of the problem include nonstationarity of EEG signals (variance for the same subject) \cite{wolpaw2002brain} \cite{vidaurre2010towards}. An additional cause, as noted by participating teams in Cybathlon 2016, is the change in the subject's emotional state. During the race, As expected, a public event such as the BCI race, the pilots' stress stress levels increased. This is to be expected as a public event such as the BCI race can heighten stress. This change in the pilots' emotional state caused their respective BCI systems to perform sub-optimally. 

The objective of this work is to mitigate this concern and develop MI systems that are robust to perturbations in the subject's emotional state, specifically to emotional arousal. In order to achieve this, we develop VR environments to induce high and low arousal in the subject before recording MI data. VR environments have been previously used along with EEG to prompt changes in emotional arousal \cite{baumgartner2006neural}. Additionally, they have been used together with MI for treating Parkinson's disease \cite{mirelman2013virtual}. To our knowledge, this is the first work where VR environments are used to increase robustness of MI-BCI systems. Subsequently, learning algorithms are trained, not only for MI but also for different arousal states. The idea is that during the BCI race, we first detect the pilot's emotional state of arousal, and choose the appropriate MI classifier. Due to COVID-19, many steps in the above mentioned outline had to be modified, the details of which are present as follows.

\section{Related Work}

\subsection{Cybathlon 2016}
The inaugural Cybathlon competition was held in 2016. After the competition, the competing teams published their methods for training the participants, amongst which were Brain Tweakers (EPFL) \cite{perdikis2018cybathlon} and Mirage91 (Graz University of Technology). One of the pilots of the former performed well in the qualifiers but poorly in the final, prompting the authors to cite psychological factors such as stress as the possible cause for the drop. A similar course of events was observed for the pilot of Mirage91, who after achieving an average runtime of 120 s in the days leading up to the Cybathlon, dropped to 196 s during the competition. The authors indicated that the pilot was showing signs of nervousness on competition day, with a heart beat of 132 beats per minute (bpm) prior to the race \cite{statthaler2017cybathlon}. 

The authors' hypothesis regarding the drop in their pilots' performances is supported by existing BCI literature \cite{chaudhary2016brain} \cite{lotte2013flaws} \cite{hammer2012psychological} \cite{jeunet2016standard}. Further support comes from evidence in affective science: It has been theorized that any event that causes an increase in emotional arousal can affect perception and memory in a manner which causes the retention of high-priority information and disregard of low-priority information \cite{mather2012selective}.

\subsection{Emotional Valence and Arousal}
\begin{figure}[tpb]
  \centering
  \includegraphics[width=0.4\textwidth]{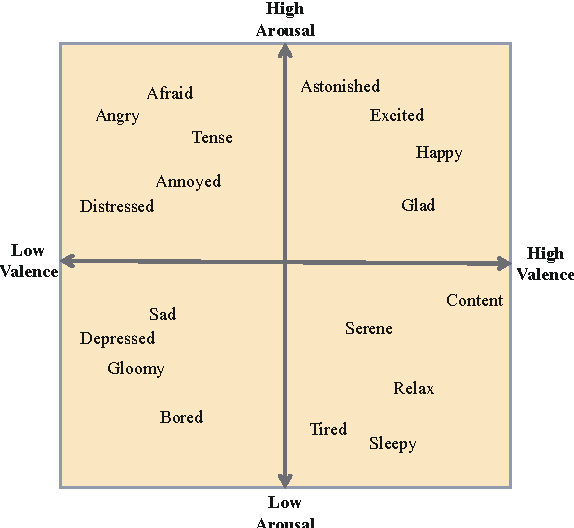}
  \caption{The circumplex model of emotional classification. Figure courtesy of \cite{Kim2016ImageRW}.} \label{fig:circumplex}
\end{figure}

Emotions are defined as complex psychological states, with three constituents: subjective experience, physiological and behavioral response \cite{hockenbury2000discovering}. Following early attempts \cite{wundt1897outlines}, more rigorous descriptions of emotions were made, the most widely accepted of which being the 'circumplex model' \cite{russell1980circumplex}. It proposes that all emotions can be described as a combination of two properties: valence and arousal. These can be thought as orthogonal axes in two-dimensions. Neurologically, it entails that any emotional state is a result of two distinct and independent neural sub-systems \cite{posner2005circumplex}. Figure \ref{fig:circumplex} provides a visual representation of the circumplex model. As can be seen, emotions such as 'excited' are high on both the arousal and valence axes, while 'gloomy' is low in both arousal and valence.

Alternative descriptions, such as the 'vector model' \cite{Bradley1992RememberingPP}, do not veer off sharply from the circumplex model; they too base emotional classification on both valence and arousal. Hence the circumplex model was used as the paradigm of emotional analysis for the duration of the project.

\subsection{Arousal, EEG and Motor Imagery}\label{arousal-eeg-background}
States of high and low arousal can be inferred from EEG signals \cite{pizzagalli2007electroencephalography}. This has been previously used to train learning systems for distinguishing between various arousal states \cite{nagy2014predicting}. EEG bands pertinent to different states of arousal are alpha (8-14 Hz) -- related to a relaxed yet awakened state -- and gamma (36-44 Hz) -- a pattern associated with increased arousal and attention. The theta pattern (4-8 Hz), correlated with lethargy and sleepiness, is also useful for differentiating arousal. 

With regards to motor imagery (MI), the most relevant EEG bands have been shown to be alpha (8-14 Hz) and beta (14-30 Hz) \cite{graimann2010brain}, the latter of which is associated with high degrees of cognitive activity \cite{pizzagalli2007electroencephalography}. 

\begin{figure}[htpb]
  \centering
  \includegraphics[width=0.3\textwidth]{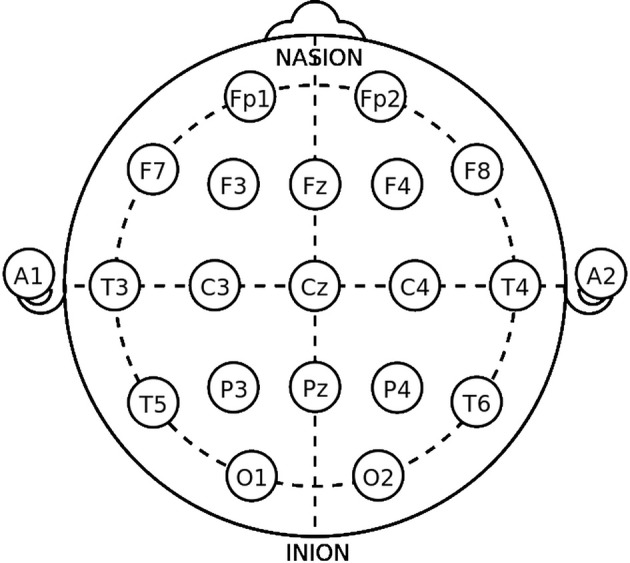}
  \caption{10-20 International system of EEG electrode placement. Electrodes C3 and C4 are most relevant for MI activity. Figure courtesy of \cite{rojas2018study}.} 
  \label{fig:eeg-map}
\end{figure}

Motor imagery data refers to data produced when the subject simulates limb movement. As movement of different limbs is sufficiently distinguishable, this can be used to perform control for various other tasks \cite{padfield2019eeg}. To record EEG data for motor imagery, the 10-20 international system of electrode placement is used \ref{fig:eeg-map}. Due to the cross-lateral nature of limb control in the human brain, movement of the right arm is recorded most faithfully by C3 and that of the left arm by C4 \cite{graimann2010brain}.

\section{Methodology}

\subsection{Virtual Reality Environments}
Traditional methods of inducing stress include the Sing-a-song stress test (SSST) \cite{brouwer2014new} and the Trier social stress test (TSST) \cite{kirschbaum1993trier}, while meditation has been shown to induce relaxation \cite{sedlmeier2012psychological}\cite{lumma2015meditation}. Emulating such environments faithfully in VR is sufficiently challenging, and may not be the most productive way to use VR to induce high/low emotional arousal.

Previously, VR exposure therapy has been explored to alleviate various psychological disorders \cite{krijn2004virtual}. One such example is using a VR height challenge -- placing the subject on higher ground in a virtual environment \cite{diemer2016fear}. Not only does the challenge induce high emotional arousal in test subjects, but the control subjects -- the ones who are not acrophobic -- also exhibit the same physiological responses as the test group i.e. increased heart rate and skin conductance level \cite{diemer2016fear}. Similarly, VR environments, particularly those with natural scenery e.g. a forest, have shown efficacy in reducing stress \cite{anderson2017relaxation}\cite{annerstedt2013inducing}. We thus developed two VR environments: one where the subject was placed on top of a skyscraper, called 'Height' while the second in a relaxing forest called 'Relaxation.' The environments were created using Unity 3D\footnote{\href{https://unity.com/}{https://unity.com/}}.

\begin{figure}[tbp]
\centering
  \begin{subfigure}[b]{0.4\textwidth}
    \includegraphics[width=\textwidth]{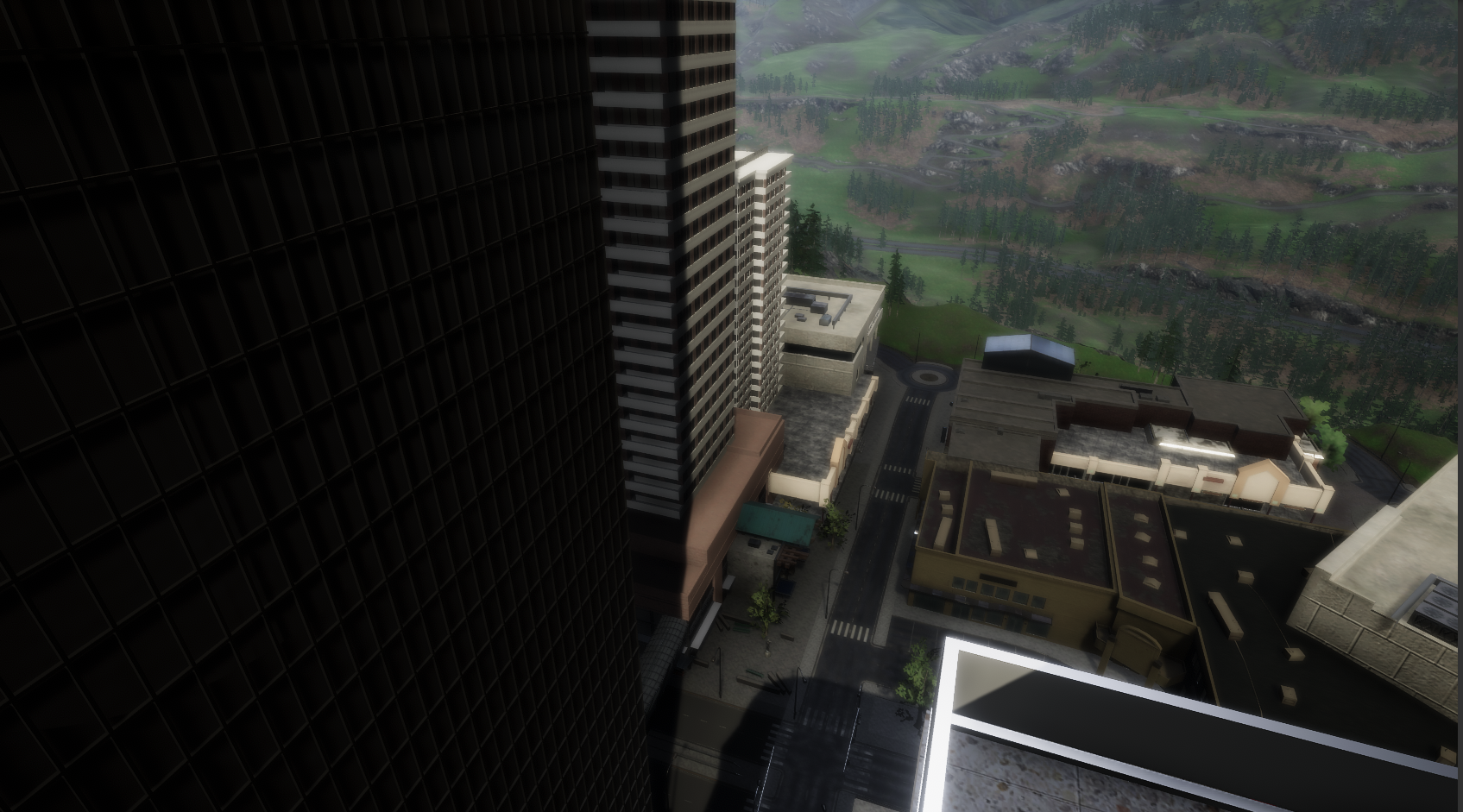}
    \label{fig:f1}
  \end{subfigure}
  \begin{subfigure}[b]{0.4\textwidth}
    \includegraphics[width=\textwidth]{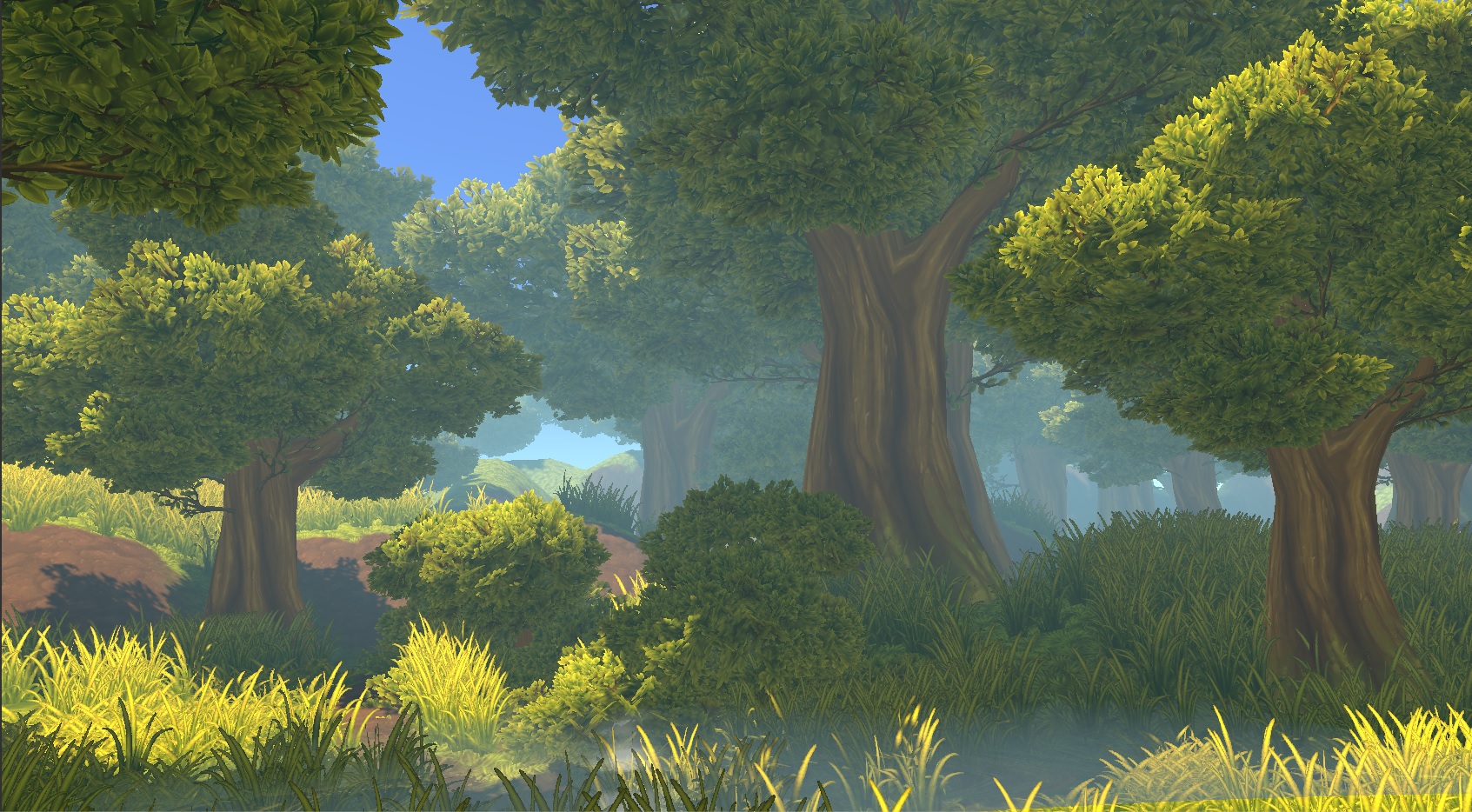}
    \label{fig:f2}
  \end{subfigure}
  \caption{Virtual reality environments for inducing arousal in subjects. On top is 'Height' designed to induce high arousal by placing the subject on the edge of a skyscraper. Below is 'Relaxation' intended to lower arousal via a natural, calming setting.}
  \label{fig:vr}
\end{figure}

\subsection{Dataset}
As this project was part of the CyberTUM team's participation in Cybathlon 2020, the original idea was to collect real data with the actual pilots who will be competing in the even proper. At the beginning of this work, however, no ethics approval had been acquired to run any experiments on the pilots. This was not detrimental to the project as a proof-of-concept could still be arrived at by collecting EEG data from volunteers within the CyberTUM team. The COVID-19 pandemic obstructed our means of collecting such data.

In the absence of our own motor-imagery and arousal data, we opted for the Graz 2b data set \cite{leeb2008bci}. It belongs to a family of BCI datasets collected by the BCI Lab at Graz University of Technology. The dataset has been used previously in the BCI Competition IV \cite{tangermann2012review}. EEG data is collected for 9 subjects doing a binary motor-imagery task (moving right and left hand on cue). The data is sampled at a frequency of 250 Hz with 3 EEG and 3 EOG channels. For our experiments, we use data from two subjects, B05 and B04, whom we refer to as subject 1 and 2 respectively henceforth.

\subsection{Subject Classification as Proxy for Arousal Classification}\label{subject-classification}
As mentioned, we were unable to obtain our own EEG arousal data. To train the classifiers, we alternatively modified the experiment. Instead of using data with high/low arousal emotional states as labels, we used different subjects as proxies for such states, making it a cross-subject classification task \cite{del2014electroencephalogram} \cite{riyad2019cross}. As EEG signals demonstrate significant variance between subjects, we can consider the data coming from subject A as that belonging to the emotional state of high arousal, and data from subject B as belonging to low arousal. With this approach, we can continue to train a classifier that would approximate the performance of one that is trained on actual arousal data, assuming the emotional states in this actual data are informative.

\subsection{Experimental Design}
The original scheme was to:
\begin{enumerate}
    \item Develop VR environments in line with existing literature that are known to induce stress (high arousal) and relaxation (low arousal) in subjects.
    \item Use electrodermal activity (EDA) activity to validate the efficacy of VR environments. EDA is a wide-used measure for emotional arousal, as skin conductance rises with rise in arousal \cite{critchley2002electrodermal}.
    \item Record MI data alternating between states of low and high arousal for each session. Start with 60s of inducing high arousal via the "Height" environment, then record MI data for 45s. Repeat the same with "Forest" environment for relaxed state. Repeat this process for each trial. The MI data was to be recorded by using the common paradigm of showing the participant a cue on screen (typically left or right arrow) which would prompt them to imagine as if they were moving their left or right hand \cite{ramoser2000optimal} \cite{pfurtscheller1997eeg} \cite{liu2017motor}.
    \item Train an arousal classifier. The aim of this classifier is to indicate the emotional state (high or low arousal) of the subject.
    \item Train separate MI classifiers for each emotional state. The goal is to optimize for accuracy, even if different types of pre-processing and classifier types were required for each state, unlike the arousal classifier which necessitates the same pre-processing steps.
    \item During deployment, first classify the emotional state using the arousal classifier, and based on its result, choose the appropriate MI classifier.

\end{enumerate}
As mentioned previously, due to numerous factors, many steps in the above formulation had to be either abandoned (2 and 3) or modified (4 and 5). The revised scheme, replaced steps 4-6 with the following:

\begin{enumerate}
    \item Train a cross-subject classifier replacing the arousal classifier. The task of this classifier is to take EEG as input from any of the two subjects, and classify the input as belonging to either subject 1 or 2. As the classifier is agnostic to the subject, the same pre-processing had to be done for each subject's data.
    \item Train separate MI classifiers for each subject instead of training for each emotional state.
    \item At test time, sample a run of a few data points (5 in our experiments), feeding them to the cross-subject classifier. Based on its mode (most frequent classification), select the appropriate MI classifier.
\end{enumerate}

\subsection{Learning algorithms}\label{algos}
We experimented with a multitude of machine learning algorithms which are briefly described as follows.

\paragraph{Logistic regression}
Logistic regression is a modification of linear regression for a binary classification task \cite{kleinbaum2002logistic}. It predicts the probability of a class given the input, by first learning a weighted linear combination of input features and applying a logistic function to the result.
\begin{equation}
    y = \frac{1}{1+e^{-a}} \quad where \quad a = \theta_0 + \theta_1.x_1 + \theta_2.x_2
\end{equation}

\paragraph{Linear discriminant analysis}
LDA attempts to maximize inter-class variance while minimizing intra-class variance \cite{balakrishnama1998linear} in the data. This results in a clustering of the data where it is easily separable. It is widely used in MI BCI \cite{wang2006common} \cite{wang2006common}.

\paragraph{Naive Bayes}
A probabilistic classifier, naive bayes uses bayes' law to calculate the posterior probability of an event (class) given the prior and likelihood \cite{murphy2006naive}. The posterior can then be updated with new evidence. It assumes that the features are independent, hence the term naive in its name.
\begin{equation}
    P(y|x) = \frac{P(y).P(x|y)}{P(x)}
\end{equation}

\paragraph{Ensemble model}
This is implemented as a voting classifier in gumpy. It uses a mix of classifiers such as nearest-neighbor, LDA and support vector machines (SVM) and uses the majority vote as the classification output. As such, it necessarily either equals or outperforms both Naive Bayes and LDA as it uses them in the ensemble.

\section{Results}

\subsection{Artifact Removal}
\begin{figure}[!t]
  \begin{subfigure}[b]{0.5\textwidth}
    \includegraphics[width=\textwidth]{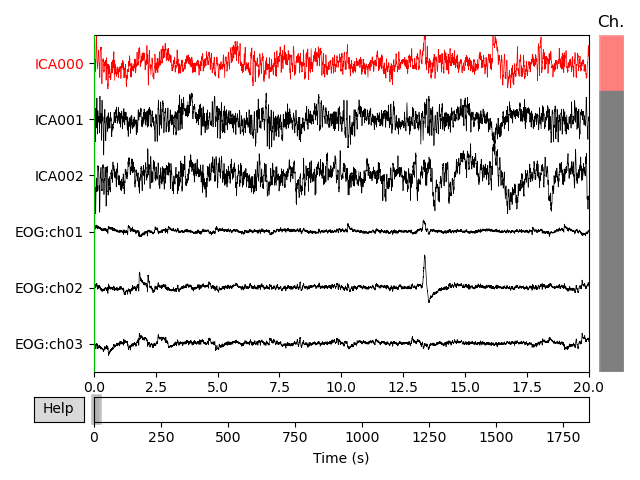}
    \caption{Plotting ICA with EOG channels. A visual depiction of the first component (in red) of ICA being correlated with EOG.}
    \label{fig:f1}
  \end{subfigure}
  \begin{subfigure}[b]{0.5\textwidth}
    \includegraphics[width=\textwidth]{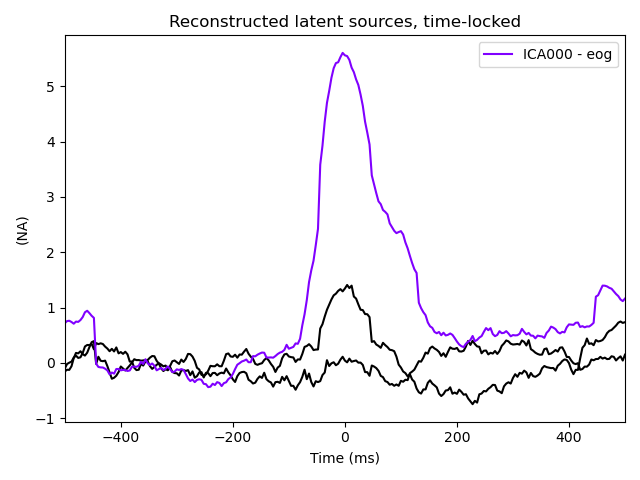}
    \caption{Plotting ICA components against each other. The peak in the first component (blue) evidently due to an eye-blink. }
    \label{fig:f2}
  \end{subfigure}
  \caption{Artifact analysis using ICA for subject 1.}
  \label{ica}
\end{figure}
The data for subject 1 and 2 contained 324 and 399 trials (attempts at moving right or left hand) respectively. The standard approach to train MI classifiers is to analyze data and remove existing artifacts before extracting features from the data \cite{uriguen2015eeg}. We first applied a Butterworth bandpass filter \cite{daud2015butterworth} to extract frequencies within the range 2-60 Hz. We then analyze the data for artifacts. A common source of artifacts in MI data is noise from electrodes located in the forehead's proximity. This is in fact data collected from the Electrooculography (EOG) channels which detect movements such as eye blinks, which may show up in the MI data. Such noise can be detected by first performing independent component analysis (ICA) -- widely used in EEG preprocessing \cite{ica} -- which tries to decompose a signal into constituent component under the assumption of statistical independence. We then see which of the resultant components correlates most with EOG channels, and filter it out \cite{ica-eog}. An example of ICA on subject 1 can be seen in figure \ref{ica}. We filter out the first component which seems to be picking up an eye blink. ICA on subject 2 did not improve the results.

\subsection{Feature Extraction}
\begin{figure*}[t]
\centering
\begin{subfigure}[b]{0.49\linewidth}
\centering
\includegraphics[height=3cm]{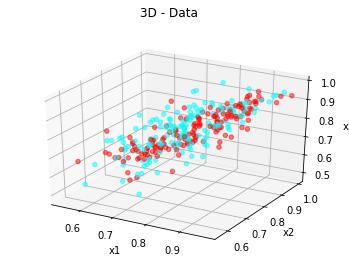}
\caption{PCA visualization of subject 1's feature vector.}
\end{subfigure} 
\begin{subfigure}[b]{0.49\linewidth}
\centering
\includegraphics[height=3cm]{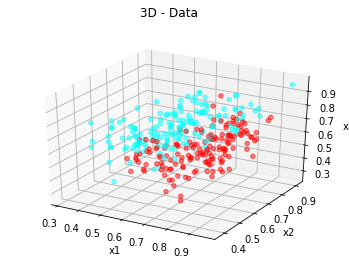}
\caption{PCA visualization of subject 2's feature vector.} 
\end{subfigure}
\caption{Dimensionality reduction using PCA for feature space visualization of both subjects. Subject 2's features are more informative for the motor-imagery task compared to subject 1 which is also reflected in the training accuracy. Right hand movements are labeled red while left hand movements are blue.}
\label{fig:pca}
\end{figure*}
Several methods were attempted to extract features. In principle, feature extraction in BCI takes two forms: frequency band selection and channel selection (also known as spatial filtering). In regards to the former, we've previously mentioned in \ref{arousal-eeg-background} that alpha and beta bands have been shown to be most related to MI activity. Accordingly, we use these frequency bands as our features. In the same section we observed that channels C3 and C4 are the most relevant for MI, which we can use directly without any spatial filtering. For this, instead of using raw alpha and beta patterns, we opt for logarithmic sub-band powers of said patterns (see gumpy documentation\footnote{\href{http://gumpy.org/}{http://gumpy.org/}}). Each spectrum is divided into four sub-bands. An alternative approach for feature extraction in MI classification has been the use of the "common spatial pattern (CSP)" algorithm \cite{Koles2005SpatialPU}. It tries to find optimal variances of subcomponents of a signal \cite{csp} with respect to a given task. In our experiments, however, CSP performed poorly compared to logarithmic sub-band power of alpha and beta bands. The results when CSP was applied have thus been omitted from the report, but could be reproduced in the notebook (see section \ref{documentation}). A visualization of the features using PCA for both subjects can be seen in figure \ref{fig:pca}. As can be observed, the features for subject 2 are more conducive to discrimination of MI. This is also verified in the training results, where every classification algorithm achieved higher accuracy for subject 2 compared to subject 1.

\subsection{Training}
\begin{figure}[t]
  \centering
  \includegraphics[width=0.5\textwidth]{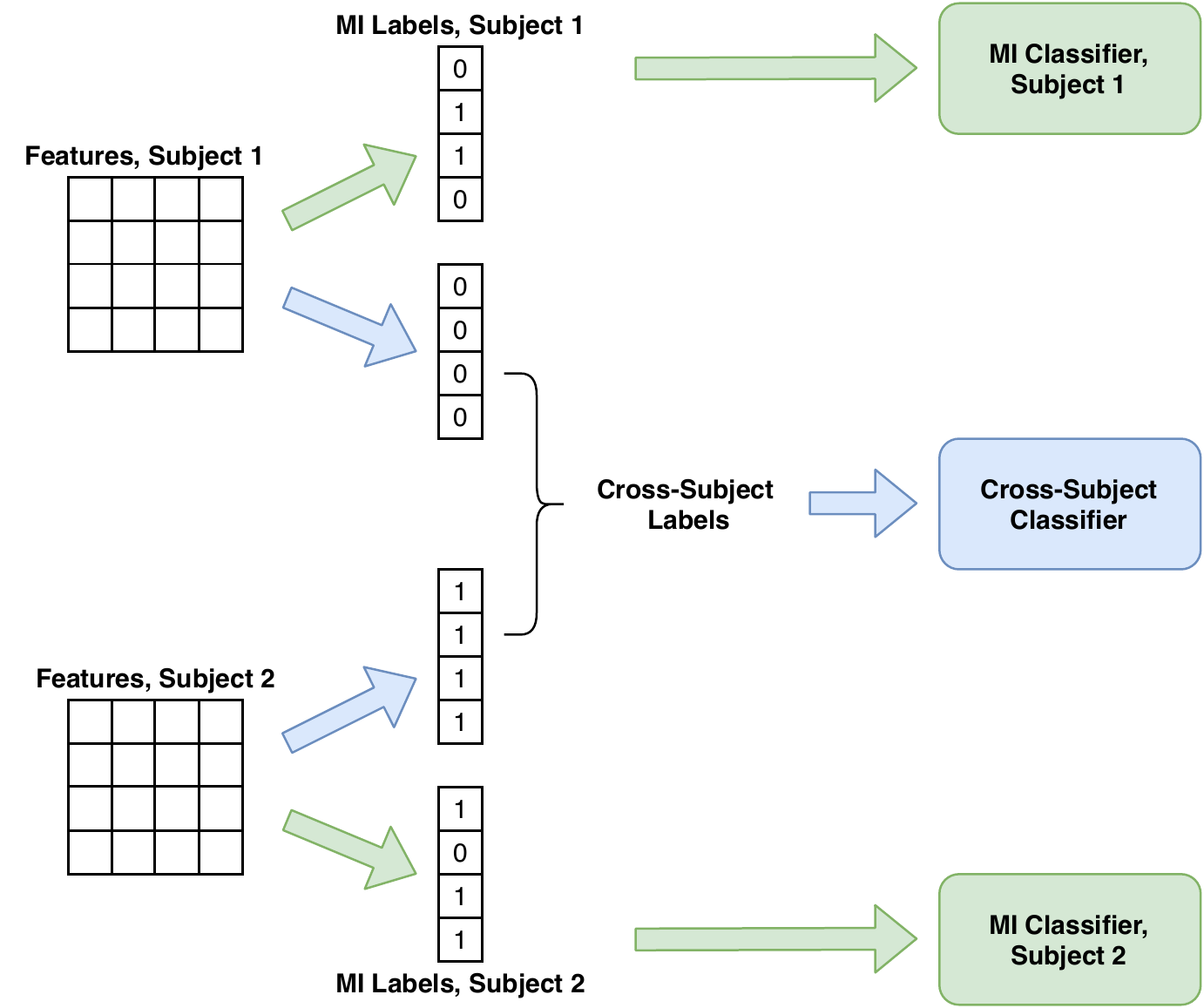}
  \caption{Training scheme for both classifiers. MI classifiers are trained separately for each subject (labels corresponding to right and left hand) while Cross-subject classifier trained on features of both subjects.} \label{fig:training}
\end{figure}
As mentioned previously, we train two types of classifiers: MI per subject classifier and cross-subject classifier. The entire training procedure is visually depicted in figure \ref{fig:training}. After doing feature extraction, we first train an MI classifier for each subject with labels 0 and 1 (left and right hand movement respectively). Subsequently, we combine data of both subjects, labelling it 0 and 1 (subject 1 and subject 2 respectively) and train the cross-subject classifier. All classifiers described in \ref{algos} are trained in each case, the results of which can be seen in table \ref{results}.

\paragraph{MI classification}
The data for each subject was divided into an 80-20 split (training-test). The features were also standardized by rescaling to zero mean and unit standard deviation. Results for both subjects were satisfactory, although subject 1's data was harder to train on compared to subject 2. This can be observed by looking at the ranges of training accuracy for both subjects [55.84-70.12 vs. 91.25-95]\%. Subject 2's classifiers achieved both a higher average accuracy as well as lower variance. LDA performed best for subject 1, while logistic regression achieved best results for subject 2.

\paragraph{Cross-subject classification}
Training for cross-subject classifiers followed the same procedure of feature extraction with the only difference being a re-labeling of the samples from limb movements to source subject. Once again, we split the data into 80-20 (train-test) portions, though this time the data is the combined samples from both subjects. For testing the classifiers, we split the test set further into sections containing five samples (trials) each. For each section, we take the mode (most frequent prediction) of the classifier which is considered the final result. For example, if our test data has 50 samples from each subject, we portion it into 20 sections (each subject with 10 sections). We then feed each section to the classifier and take the majority score for that section as the classifier's prediction. As can be seen, the ensemble model outperforms the rest of the algorithms by a considerable margin. In addition to this, we also created t-SNE embeddings of the features with 2 and 3 dimensions \cite{tsne}. The results were not up to par and have thus been left out here (they can be reproduced via the notebook discussed in \ref{documentation}). More details can be found in \ref{discussion}.

\begin{table}[h]
\caption{Summary of results. Accuracy scores for MI (both subjects) as well as cross-subject (X-sub) using various classifiers. Best results in bold}
\begin{center}
\begin{tabular}{|c|c|c|c|c|}
\hline
\textbf{Task}  &   \multicolumn{4}{|c|}{\textbf{Classifier}} \\
\cline{2-5} 
\textbf{} &  \textbf{Logistic Regression} & \textbf{LDA} & \textbf{Naive Bayes} & \textbf{Ensemble}\\
\hline
MI-sub 1    &   67.53\%    &   \textbf{70.12\%}    &     55.84\%     &   70.12\% \\
\hline
MI-sub 2    &   \textbf{95\%}    &   91.25\%    &    91.25\%     &   93.75\% \\
\hline
X-sub    &   58.65\%    &   59.38\%    &    59.38\%     &   \textbf{68.75\%} \\
\hline
\end{tabular}
\label{results}
\end{center}
\end{table}

\section{Discussion}\label{discussion}

The results indicate that assuming different emotional states impart sufficient differences in EEG data, we can train classifiers that perform well above chance. Significant differences in the EEG signals between both subjects were observed during feature extraction and classification. This is not an uncommon phenomenon and has been documented in the literature \cite{pizzagalli2007electroencephalography}. Blankertz et. al show that after testing on 80 subjects, the average classifier accuracy of a binary task was 74.4\% with a spread of 16.5\% \cite{BLANKERTZ20101303}. Our findings buttress this as the best models for subject 1 and 2 achieved 75.38\% and 95\% accuracy respectively. This variability generally chalked up to differences in the subjects' abilities for implicit learning \cite{kotchoubey2000learning}, performance in early neurofeedback sessions \cite{Neumann1117} and attention spans \cite{Daum94}. 

According to Tangermann et. al, the best results on data set 2b were achieved using filter-bank CSP as a pre-processing step followed by a naive bayes classifier during Competition IV \cite{tangermann2012review}. In our testing, however, vanilla CSP for feature extraction was sub-optimal. Naive bayes was also found trailing behind other classifiers as seen in table \ref{results}. We thus observe that vanilla CSP is not as performant as log band-power in our experiments, while we did not perform any experiments with filter-bank CSP.

In regards to cross-subject classification, appreciable results have been achieved by using ICA for feature extraction \cite{tangkraingkij2009selecting} combined with a nearest-neighbor (NN) classifier. We verify the efficacy of ICA as a pre-processing step for feature extraction. Other approaches have shown PCA as an effective step for dimensionality reduction \cite{palaniappan2005energy}. While we could not confirm this with PCA, using the more modern dimensionality reduction technique of t-SNE performed poorly in our experiments (tested using target dimensions 2 and 3). There is, however, recent evidence that using t-SNE in tandem with common dictionary learning may yield good results \cite{nishimoto2020eeg}.

\subsection{Limitations and Future Outlook}

A primary limitation of this work is the lack of testing on actual subjects. While the system ensures acceptable performance on an existing dataset, we can not conclude much about its usefulness in the real-world. To make such assertions with a certain degree of confidence, we need to evaluate how quickly we can switch between various MI classifiers based on the predictions of the emotion (cross-subject) classifier. This is also true for calibration time at the start of each session; while we use five trials during testing and get well above-chance results, comprehensive and systematic verification of the system is in order if it is to be of any practical use.

In addition to alpha patterns, gamma bands are correlated with increased arousal \cite{pizzagalli2007electroencephalography}, which may have carried a strong supervision signal for the classifier. Had we acquired EEG data for aroused and relaxed states of a subject, an emphasis on gamma bands would have been warranted. As such, in the present case, as we did not have data corresponding to high and low arousal, gamma patterns were assumed not to be informative.

Future work may also look at training classifiers for more than two subjects. While two subjects suffice for the purposes of this study, as the original task was the discrimination between two emotional states of arousal, it may be worth exploring how the cross-subject classifier would scale to additional classes. This may be interpreted as having to classify not only emotional arousal but also valence (positive or negative) which may have important ethical implications.

Most of the classifiers used in this project are classic algorithms, and were chosen for their still prevailing use in MI BCI. However, future work may also incorporate modern approaches such as deep neural networks for MI classification \cite{Tabar_2016}. Deep learning could also be used to formulate our problem as that of multi-task learning for both arousal and MI classification \cite{multitask}. In this manner we can replace training multiple classifiers with a single one which both classifies emotional arousal as well as motor imagery.

\section{Interdisciplinary Work}
The nature of this project necessitated the undertaking of a multi-disciplinary approach, from understanding and systematizing human emotional arousal to developing algorithms for distinguishing both emotional states as well as motor function via EEG. Thus, this work borrows, incorporates and synthesizes elements from a number of disciplines including psychology (emotional arousal), neuroscience (EEG and motor-imagery), computer graphics (virtual reality environments) and artificial intelligence (machine learning for classification). Broadly, we can categorize psychology and neuroscience as brain sciences and computer graphics and artificial intelligence under the umbrella of informatics. Each of the two disciplines contributed unique methods and insights without which the project may not have come to fruition. The most valuable insight was the difficulty in training accurate machine learning algorithms for EEG. Although machine learning has become the dominant paradigm for classification tasks, this project demonstrates that pre-processing of data (via techniques such as ICA and log power-band) is at least as important to the success of the system as the classifier (the results for other feature extractors can be reproduced in the provided notebook), and even after pre-processing, we have no guarantees of robust performance. Another key insight was the extent to which EEG patterns vary between different people, pointing to the difficulty of transfer learning in this domain.

\section{Conclusion}
A major hurdle in the widespread and practical use of assistive systems based on MI-BCI is lack of reliability. While this can have many origins, an important source as identified by two Cybathlon teams in 2016 was related to shifts in the subject's state of emotional arousal. In this work, we present an end-to-end framework for inducing high/low arousal in subjects, collecting EEG data and train learning algorithms for robust MI classification. While COVID-19 enforced certain constraints on data acquisition, we were still able to develop a proof-of-concept for how emotion-robust MI-BCI systems could be trained. Our results indicate that if the training signal contains sufficient information i.e. each emotional state has a distinct enough EEG signature, we can successfully train systems that are robust to variance in emotional arousal. A thorough study, however, needs to be conducted to determine the practicality of such a system with respect to variables such as classifier switching times and calibration periods.

\section{Acknowledgments}
This project could not have been possible without the aid of Nicholas Berberich who provided constant and quality guidance on overall methodology, feature extraction and algorithms. Also worth gratitude are Matthijs Pals for his support in regards to MI data preprocessing and Svea Meyer for helping in the initial phase of the project as well as with explaining EEG terminology.

\bibliographystyle{plain}
\bibliography{bibliography.bib}
\vspace{12pt}

\section{Appendices}
\subsection{Documentation}\label{documentation}
The VR environments developed for this project can be found on the following Google Drive links:
\begin{itemize}
    \item Forest VR - \href{https://tinyurl.com/y79dxk87}{https://tinyurl.com/y79dxk87}
    \item Height VR - \href{https://tinyurl.com/yaqrmveu}{https://tinyurl.com/yaqrmveu}
\end{itemize}
The environments were created using Unity version 2018.4.14f with the post-processing stack enabled.
All code written for this project pertaining to training can be found on LRZ Gitlab using the following link: \href{https://gitlab.lrz.de/cybertum/eeg-classification}{https://gitlab.lrz.de/cybertum/eeg-classification} and \href{https://github.com/Abdul-Moeed/eeg-classification}{https://github.com/Abdul-Moeed/eeg-classification}. The Jupyter notebook titled 'Classifier Demo' contains the crux of the code; the rest of the scripts were created for testing tools and techniques. To reproduce ICA plots, run 'classifier.py'. The data used in this project can be found at \href{http://www.bbci.de/competition/iv/}{http://www.bbci.de/competition/iv/} under 'Data sets 2b'. Further information on reproducibility can be found in the repository's README.

\subsection{Methodological Reflections}

\begin{enumerate}
  \item \textit{What would you do differently next time in your experimental/technical setup?} A major challenge, compounded by COVID-19, was the collection of arousal data. We were originally planning to use the Biopac system (BIOPAC Systems, Inc., Goleta, California, United States) for measuring skin conductance. The procedure is involved and somewhat cumbersome, and could have been substituted with a heart rate sensor. Heart rate measurement is also an indicator of emotional arousal levels, and has the added benefit of being present in a smart watch. This would have made data acquisition simpler. Another important step would have been to test the efficacy of VR environments in inducing high/low arousal in systematic trials under control. Although there already exists evidence that VR has been known to impart such states, to observe its effect on EEG data would have been desirable. Finally, while classic machine learning is still widely used in MI classification, we observed, at least in subject 1's case, the difficulty in training a robust classifier. Perhaps opting for deep learning, which has shown increased performance in other domains such as computer vision and natural language processing, may have been promising.
  
  \item \textit{What went really good/easier than expected? } The creation of VR environments was surprisingly simple due to vast community support for Unity 3D. Integrating VR controls in the environments was slightly more involved but went rather smoothly. Regarding training: Besides feature extraction, training the classifiers was fairly straightforward. This was in part due to previous experience working with machine learning and in part because of easy-to-use implementations of common algorithms in existing libraries.
  
  \item \textit{What kinds of skills would you need to learn in your studies to do such a project even better (machine learning, statistics, psychology, neuroscience)?}
  Taking a course on BCI prior to working on this project would have substantially accelerated development. Trying to get familiarized with methods and terminology of the field was a major part of the project, but it also entailed allocating time for such self-study which could have been dedicated to development had prior knowledge in this domain been acquired. Thus a course on neurofeedback and BCI would have made the project better.
  
  \item \textit{If you tried out an approach which you later on abandoned, you can report this here as well. Give reasons why you decided not to pursue this approach.} During the training of MI classifiers, initially I used data from certain sessions for each subject. This yielded scores with high variance, even for the same subject for different sessions. I abandoned this by taking data from all sessions for each subject and using that for training instead. 
  
\end{enumerate}

\end{document}